\documentclass[12pt]{article}
\usepackage{amsmath,amssymb,amsthm,amsfonts,bm,mathrsfs,float,algorithm,algorithmic,caption,subcaption,setspace,longtable,tabularray}
\usepackage{avant,array,geometry,inputenc,natbib,hyperref,multirow,enumerate,enumitem,rotating,booktabs,color,xcolor,latexsym,stmaryrd,xr}
\usepackage[T1]{fontenc}
\usepackage[hang,flushmargin]{footmisc}

\usepackage{blindtext, soul}
\usepackage[normalem]{ulem}
\usepackage{amsmath,amssymb,amsthm,amsfonts,bm,mathrsfs,float,algorithm,algorithmic,caption,subcaption,setspace}
\usepackage{avant,array,multirow,enumerate,enumitem,rotating,color,latexsym,stmaryrd,xr,tikz,algorithm,amsmath,amsfonts,mathrsfs}
\SetSymbolFont{stmry}{bold}{U}{stmry}{m}{n}
\usepackage[mathscr]{euscript}
\usepackage{enumitem}
\usepackage{graphicx}

\makeatletter
\newcommand{\mylabel}[2]{#2\def\@currentlabel{#2}\label{#1}}
\makeatother

\numberwithin{equation}{section}

\newtheorem{remark}{Remark}
\newtheorem{theorem}{Theorem}
\newtheorem{corollary}{Corollary}

\def\A{\boldsymbol{\mathrm{A}}}

\def\D{\boldsymbol{\mathrm{D}}}

\def\H{\boldsymbol{\mathrm{H}}}

\def\I{\boldsymbol{\mathrm{I}}}

\def\P{\boldsymbol{\mathrm{P}}}
\def\R{\boldsymbol{\mathbf{R}}}

\def\Y{\boldsymbol{\mathrm{Y}}}

\def\Q{\boldsymbol{\mathrm{Q}}}

\def\0{\boldsymbol{0}}
\def\1{\boldsymbol{1}}

\def\k{k_*}
\def\mk{m_{k_*}}

\def\bSigma{\boldsymbol{\Sigma}}
\def\bOmega{\boldsymbol{\Omega}}

\def\betaa{\boldsymbol{\eta}}
\def\bdelta{\boldsymbol{\delta}}

\def\bR{\mathbb{R}}

\def\bK{\mathbb{K}}

\def\cH{\mathcal{H}}
\def\cI{\mathcal{I}}

\def\cS{\mathcal{S}}

\def\scY{\mathscr{Y}}

\def\scZ{\mathscr{Z}}
\def\scM{\mathscr{M}}
\def\scV{\mathscr{V}}

\def\bscY{\boldsymbol{\mathscr{Y}}}

\def\bscZ{\boldsymbol{\mathscr{Z}}}

\def\bscM{\boldsymbol{\mathscr{M}}}
\def\bscV{\boldsymbol{\mathscr{V}}}

\def\TN{\mathrm{Tnorm}}
\def\VN{\mathrm{mnorm}}

\def\trans{^{\scriptscriptstyle \sf T}}
 
\def\vec{\mathsf{vec}} 

\def\id{\mathbb{I}}  
\def\ep{\mathsf{E}} 
\def\Cov{\mathsf{Cov}} 
\def\Var{\mathsf{Var}}

\def\inv{^{-1}}

\newcommand{\bs}{\backslash}

\begin{document} 

\title{Alteration Detection of Tensor Dependence Structure via Sparsity-Exploited Reranking Algorithm}

\author{
\bigskip
Li Ma$^\dag$, Shenghao Qin$^\dag$, and Yin Xia$^\dag$ \\
\normalsize{\textit{$^\dag$Department of Statistics and Data Science, Fudan University}}
}

\date{}

\maketitle 
\begin{abstract}
Tensor-valued data arise frequently from a wide variety of scientific applications, and many among them can be translated into an alteration detection problem of tensor dependence structures. 
In this article, we formulate the problem under the popularly adopted tensor-normal distributions and aim at two-sample correlation/partial correlation comparisons of tensor-valued observations. 
Through decorrelation and centralization, {a separable} covariance structure is employed to pool sample information from different tensor {modes} to enhance the power of the test.
Additionally, we propose a novel \textbf{S}parsity-\textbf{E}xploited \textbf{R}eranking \textbf{A}lgorithm (SERA) to further improve the multiple testing efficiency. 
The algorithm is approached through reranking of the $p$-values derived from the primary test statistics, by incorporating a carefully constructed auxiliary tensor sequence. 
Besides the tensor framework, SERA is also generally applicable to a wide range of two-sample large-scale inference problems with sparsity structures, and is of independent interest.
The asymptotic properties of the proposed test {are} derived and the algorithm is shown to control the false discovery at the pre-specified level.
We demonstrate the efficacy of the proposed method through intensive simulations and two scientific applications. 
\end{abstract}


\newpage
\baselineskip=22pt

\section{Introduction}
The growing accessibility of multi-dimensional data has attracted increasing attention to tensor inference. Tensors are higher-order parallels of vectors (first-order) and matrices (second-order), and are denoted by bold Euler letters such as $\bscY$ throughout the paper. {Specifically, a $K$-th order tensor with dimension vector $(m_1,\dots,m_K)$ is denote by $\bscY\in\bR^{m_1\times m_2\times\cdots\times m_K}$, and it has $K$ ``{modes}'' in total} \citep{hoff2011separable}. 
As a concrete example, the international trade dataset in Section \ref{subsec:real-data-corr} collects monthly imports of 97 commodity types over 30 countries from the year 2015 to 2022. The data from each year is a {$97\times 30\times12$} tensor-valued observation, where the commodity types, countries and months serve as the first, second and third mode, respectively. 
Besides, tensor-valued data also arise in many other fields, including climate change detection (e.g., the analysis in Section \ref{subsec:real-data-pcorr}), 
{gene microarray study \citep{zahn2007agemap,hore2016tensor}, neuroimaging research \citep{stolp2018voxel, zhou2023partially}, 
recommendation system analysis \citep{wang2019tensor,zhang2021dynamic}, among many others.}

This article targets the dependence structures of tensors, and the samples are assumed to follow some tensor normal distributions with a separable covariance matrix. Specifically, we assume that a $K$-th order tensor $\bscY\in\bR^{m_1\times m_2\times\cdots \times m_K}
\sim \TN(\bscM,\bSigma_1\circ\cdots\circ\bSigma_K)$, where ``$\TN$'' stands for ``tensor normal'', {$\bscM$ and $\bSigma_1\circ\cdots\circ\bSigma_K$ respectively represent the tensor mean and covariance}, and ``$\circ$'' is the outer product \citep{hoff2011separable}. 
Note that, the covariance matrix $\bSigma_{k}$, which is associated with each of the modes, characterizes the \emph{within-mode} dependence structure for $k=1,\ldots, K$, where $K$ is assumed to be fixed. 
The distribution with such separable covariance structure links to the multivariate normal distribution via $\vec(\bscY_{})\sim \VN(\vec(\bscM), \bSigma_{K}\otimes\cdots\otimes\bSigma_{1})$, where $\vec(\cdot)$ denotes the vectorization, $\VN(\cdot,\cdot)$ represents a multivariate (vector) normal distribution and $\otimes$ is the Kronecker product. 
Hence, such a model reduces the ultra-high dimension of the covariance from $\prod_{k=1}^Km_k^2 $ to $\sum_{k=1}^K m_k^2$. 
The verification of the covariance separability is well studied in the literatures \citep[e.g.,][]{aston2017tests,constantinou2017testing,bagchi2020test}.
In addition, such tensor-normal distribution (with $K\geq 3$) has been widely adopted in many inference problems, see for example, {\cite{hoff2015multilinear}, \cite{li2017parsimonious} and \cite{pan2018covariate}}.
When $K=2$, it reduces to matrix normal assumption with separable covariance structure, which appears frequently in the literatures as well {\citep[][among others]{efron2009set,  leng2012sparse,zhou2014gemini}}. 



\subsection{Problem Formulation and Algorithm Sketch}\label{subsec:problem}
In this article, we focus on the tensor scenarios when $K\geq 3$,
and we aim at the alteration detection between {mode-$\k$} dependence structures of the tensor-valued observations from two different groups, where $\k\in\{1, \ldots, K\}$ is a pre-specified mode of interest and the rest of modes are treated as {nuisances}.
Specifically, suppose we observe two groups of independent and identically distributed (i.i.d.) samples: $\{\bscY_{l,1}\}_{l=1}^{n_1}$ and $\{\bscY_{l,2}\}_{l=1}^{n_2}$, where
\begin{align}
	\bscY_{l,d}\in\bR^{m_1\times\cdots\times m_K}\stackrel{i.i.d.}{\sim} \TN(\bscM_d,\bSigma_{1,d}\circ\cdots\circ\bSigma_{K,d}),\; l=1,\ldots,n_d,\; d=1,2.
	\label{eqn:tensor-normal}
\end{align}
In such a model, the covariance matrices are only identifiable up to a constant, namely, $\bscY_{l,d}{\sim} \TN(\bscM_d,(c_{1,d}\bSigma_{1,d})\circ\cdots\circ(c_{K,d}\bSigma_{K,d}))$ holds as long as $\prod_{k=1}^K c_{k,d}=1$. 
Hence, instead of covariance and partial covariance, throughout we target the \emph{correlation} matrices $\R_{\k,d}^{\scriptscriptstyle (1)}=\D_{\bSigma_{\k,d}}^{-1/2}\bSigma_{\k,d}\D_{\bSigma_{\k,d}}^{-1/2}$ and the \emph{partial correlation} matrices $\R_{\k,d}^{\scriptscriptstyle (2)}=\D_{\bOmega_{\k,d}}^{-1/2}\bOmega_{\k,d}\D_{\bOmega_{\k,d}}^{-1/2}$, where $\D_{\A}$ is the diagonal matrix of $\A$ and  $\bOmega_{\k,d}=\bSigma_{\k,d}^{-1}$. 
To unify our analysis, we denote the dependence structures of interest by $\P_{\k,d} = (\rho_{i,j,d})_{\mk \times \mk}$, which can be either $\R_{\k,d}^{\scriptscriptstyle (1)}$ or $\R_{\k,d}^{\scriptscriptstyle (2)}$. Then, one wishes to detect the changes of such dependence structures between the two groups with the control of false discovery. That is, the goal is {to test} simultaneously 
\begin{align}
	H_{0,i,j}:\rho_{i,j,1}=\rho_{i,j,2} \quad \mbox{v.s.}\quad  H_{1,i,j}:\rho_{i,j,1}\neq \rho_{i,j,2}, \qquad 1\leq i< j\leq \mk,
	\label{eqn:hypothesis}
\end{align}
with false discovery rate (FDR) and false discovery proportion (FDP) control, 
and we will carry out the analysis in the following three steps. 

The first step is \emph{sample transformation}. It decorrelates and centers the original tensor-valued observations, and then utilizes the nuisance modes  to enhance the inference on mode-$\k$ dependence structure. This step is achieved by plugging in all nuisance covariance matrices and applying a careful orthogonal rotation. 
Thanks to the separable covariance structure under Model \eqref{eqn:tensor-normal}, 
by pooling the information from the nuisance modes, the effective sample size increases significantly. This step is crucial for the subsequent analysis and more explanations can be found in Section \ref{subsec:trans-sample}.

The second step is 
\emph{statistic-pair construction}.
To be specific, for each $(i,j)\in \cH=\{(i,j):1\leq i<j\leq \mk\}$, a primary $t$-statistic $T_{i,j}$ is first calculated to quantify the signal strength $\rho_{i,j,1}-\rho_{i,j,2}$. Then, an auxiliary covariate $U_{i,j}$ based on some weighted sum $\rho_{i,j,1}+\kappa_{i,j}\rho_{i,j,2}$ is constructed to capture the sparsity information. 
Note that, it is often the case that the dependence structure for each group is individually sparse under the high-dimensional setting. Hence, a signal is likely to have a large $|U_{i,j}|$ which differentiates itself from the nulls with $\rho_{i,j,1}=\rho_{i,j,2}= 0$. 
Therefore, the auxiliary statistic $U_{i,j}$ reflects the heterogeneity of the testing units and can be employed to construct more efficient testing procedures.
It is worthwhile to note that, a careful selection of $\kappa_{i,j}$ in the auxiliary statistic is necessary in order to guarantee the validity of the testing algorithm; the details will be provided in Sections \ref{subsec:stat-pair-construct} and \ref{subsec:asymp-stat}.

As the final step, we propose a novel \emph{{S}parsity-{E}xploited {R}eranking {A}lgorithm (SERA)}. 
Instead of directly applying BH procedure \citep{benjamini1995controlling} to the primary sequence $\{T_{i,j}\}$, we employ the earlier constructed auxiliary statistics $\{U_{i,j}\}$ to extract the sparsity knowledge and rerank the $p$-values.
Specifically, a set of locally adaptive weights is constructed based on $\{U_{i,j}\}$ and we place differential weights on each of the $p$-values obtained through $\{T_{i,j}\}$.
Subsequently, the false discovery estimates can be adjusted and the testing efficiency can be further improved. 


\subsection{Related Works and Our Contributions}\label{subsec:literature}
In the literature, there have been a good number of methods proposed to estimate the tensor dependence structures.
For example,  matrix-valued dependence estimations are studied in  \cite{leng2012sparse, yin2012model, zhou2014gemini, zhu2018multiple}; {\cite{hornstein2018joint,zhang2022covariance}}, among many others, and are further extended to tensor cases, such as  
tensor covariance estimations  \citep{hoff2011separable, singull2012more,manceur2013maximum,nzabanita2015maximum} and tensor graphical model estimations \citep{tsiligkaridis2013convergence,he2014graphical, xu2017efficient, min2022fast}.  
However, they mostly target {one-sample} dependence structure recovery and focus on estimation instead of testing, 
and hence cannot be directly employed to solve our problem \eqref{eqn:hypothesis}.

The multiple testing problem \eqref{eqn:hypothesis} for vector-valued or matrix-valued observations (i.e., $K=1$ or $2$) have been well studied in the literature. {Specifically, for vector-valued samples, \cite{xia2017testing}, \cite{cai2016large} and \cite{xia2015testing} study the simultaneous inference of two-sample covariance/correlation/precision matrices, respectively;
for matrix-valued cases, \cite{chen2019graph} targets  one-sample precision inference while \cite{chen2021testing} and \cite{xia2019matrix} perform two-sample correlation/partial correlation comparisons respectively; see Table 12 in the supplement of \cite{chen2021testing} for a summary of related literatures.} When $K\geq 3$, \cite{lyu2019tensor} considers {one-sample} inference on the precision matrix of a specific mode but their test cannot be directly extended to the two-sample cases with theoretical guarantees. 
As will be seen in Section \ref{subsec:simu-fdr}, a modified two-sample test based on \cite{lyu2019tensor} often performs poorly.


Our proposal differs from existing solutions and makes several useful contributions.
First in terms of problem formulation, to the best of our knowledge, there is no existing work that solves \eqref{eqn:hypothesis} under the current tensor framework.  
Hence, we fill an important gap in two-sample multiple testing of tensor correlation/partial correlation matrices.
To achieve this, we apply a new orthogonal rotation to the tensor-valued observations and take advantage of the separable covariance structures to pool information from the rest $K-1$ nuisance modes.
Through the data pooling, the effective sample size grows from $n_d$ to $n_d\prod_{k\neq\k}m_k$ which significantly improves the power of the subsequent testing procedure.
Second and methodologically, we develop a novel multiple testing method, SERA, for additional testing efficiency improvement.
Such power enhancement is approached through the $p$-value weighting scheme, where the weights are calculated based on a carefully constructed sequence of auxiliary tensor statistics.
Besides our tensor dependence testing scenarios, SERA is also generally applicable to many other vector/matrix/tensor two-sample multiple testing problems; see the detailed discussions in Section \ref{subsec:sera-indep}. 
In comparison to the existing multiple testing methods with auxiliary covariates \citep[e.g.,][]{liu2014incorporation,tony2019covariate,xia2019gap}, the proposed SERA explores the underlying sparsity structure in a continuous fashion, and enjoys power superiority, dependency robustness and computation efficiency in the meantime; see the detailed explanations and numerical comparisons in Section \ref{sec:simu}. Hence, our proposal makes a useful addition to the general toolbox of multiple testing with side information. 
Third and technically, we establish within-mode sample covariance estimation consistency through correlated fibers and derive an overall error bound for the Kronecker product of $K-1$ nuisance estimates.
Moreover, we show the consistency of a Nadaraya-Watson-type kernel estimator \citep{nadaraya1964estimating,watson1964smooth} under a random design, which is to our knowledge not available in the literature.
In addition, we extend existing simultaneous error control theories to a new setting with an additional continuous auxiliary sequence, and develop a new set of theoretical tools.
\subsection{Organization of the Paper}
The rest of the paper is organized as follows. Section \ref{sec:three-steps} studies the implementation of the proposed algorithm. Section \ref{sec:theory} collects the theoretical properties. Simulations and real data analysis are provided in Sections \ref{sec:simu} and \ref{sec:real-data} respectively. We relegate some methodological details, all technical proofs, and additional numerical informations to the Online Appendix.

\section{Three-Step Inference of T-SERA}\label{sec:three-steps}
This section studies the implementation of the proposed Tensor-valued {S}parsity-{E}xploited {R}eranking {A}lgorithm, and we name it T-SERA in short.
We first introduce some notation and tensor operations in Section \ref{subsec:preliminary}, and then delve into the three-step T-SERA in Sections \ref{subsec:trans-sample} to \ref{subsec:sera-alg}. 

\subsection{Preliminaries}\label{subsec:preliminary}

Note that, a $K$-th order tensor $\bscY$ reduces to a vector when $K=1$ and a matrix when $K=2$. 
Denote by $[n] = \{1, 2, \ldots , n\}$ for a positive integer $n$. 
Then the elements of $\bscY$ can be listed as $\{\scY_{i_1,\ldots,i_K}: i_k \in [m_k],\ k \in [K]\}$, with a total number $m=\prod_{k=1}^Km_k$. 
``{Fibers}'' of a tensor refer to the high-order analogs of matrix rows and columns, and are obtained by fixing all but one of the indices of the tensor \citep{kolda2006multilinear}. Specifically, any mode-$k$ fiber is a vector of length $m_k$ that is given by $\bscY_{i_1,\ldots,i_{k-1},:,i_{k+1},\ldots,i_K}$. {For example, a matrix column is a mode-1 fiber and a matrix row is a mode-2 fiber.} Hereinafter we denote a mode-$k$ fiber by $\bscY_{\{i_j\}_{j\in[K]\bs k}}$ for ${i_j\in [m_j]}, j\in[K]\bs k$. Furthermore, we denote the $i$-th entry of $\bscY_{\{i_j\}_{j\in[K]\bs k}}$ by $\scY_{i,\{i_j\}_{j\in[K]\bs k}}$, and denote the sub-fiber with the $i$-th entry removed by $\bscY_{-i,\{i_j\}_{j\in[K]\bs k}}$. 

Next, we briefly review the {$k$-mode} product of a tensor $\bscY\in\bR^{m_1\times m_2\times\cdots \times m_K}$ and a matrix $\A\in\bR^{J\times m_k}$ \citep{de2000multilinear}. It is denoted by $\bscY\times_k \A$, which produces a new $K$-th order tensor $\bscV\in\bR^{m_1\times\cdots m_{k-1}\times J\times m_{k+1}\times\cdots\times m_K}$ with entries
\begin{align*}
	\scV_{i_1,\dots,i_{k-1},j,i_{k+1},\dots,i_K}=\sum_{i_k=1}^{m_k}\scY_{i_1,\dots,i_K}\mathrm{A}_{j,i_k},\quad j \in [J].
\end{align*}
Then for a list of matrices $\{\A_1,\cdots,\A_K\}$ with $\A_k\in\bR^{n_k\times m_k}$ for $k\in[K]$, the Tucker product \citep{tucker1966some, kolda2006multilinear} of $\bscY\in\bR^{m_1\times m_2\times\cdots \times m_K}$ and $\{\A_1,\cdots,\A_K\}$ is defined as,
\begin{align*}
	\bscY\times\{\A_1,\cdots,\A_K\}=\bscY\times_1 \A_1\times_2 \A_2\times\cdots\times_K \A_K,
\end{align*}
which yields a $n_1\times\cdots\times n_K$ tensor.

\subsection{Step 1: Sample Transformation}\label{subsec:trans-sample}
We first explain the transformation idea for the oracle case where all nuisance covariance matrices are known. Then we provide a fully data-driven implementation in Algorithm \ref{alg:sample-trans}.

For $d=1,2$, we first stack $n_d$ observed tensors into a $(K+1)$-th order tensor denoted by $\bscY_d=(\bscY_{1,d},\cdots,\bscY_{n_d,d})$. {It is easy to check that $\bscY_d\sim \TN(\bscM_d\circ \1_{n_d},\bSigma_{1,d}\circ\cdots\circ\bSigma_{K,d}\circ \I_{n_d})$ by Proposition \ref{pro:tensor_stack} in the Online Appendix, {where $\1_{n_d} \in \bR^{n_d}$ is a vector repeating $1$'s and $\I_{n_d} \in \bR^{n_d\times n_d}$ is an identity matrix.} The proposed transformation has two goals: \emph{decorrelation} and \emph{centralization}, and can be achieved in a one-step operation. Decorrelation aims to make all mode-$\k$ fibers independent and have identical covariance matrices; it can be accomplished by plugging in all true nuisance covariance matrices. For centralization, we rotate the tensor-valued observations by {any} orthogonal matrices $\{\Q_d\in\bR^{n_d\times n_d}:d=1,2\}$ with the last row equal to {$({1}/{\sqrt{n_d}},\dots, {1}/{\sqrt{n_d}})$}; it transfers the means $\{\bscM_d:d=1,2\}$ to zero tensors. Specifically, we transform the original samples by the following Tucker product: 
\begin{align*}
	\bscZ^o_{d}=\bscY_{d}\times\{ {\bSigma}_{1,d}^{-1/2},\cdots,\bSigma_{\k-1,d}^{-1/2},\boldsymbol{\mathrm{I}}_{\mk},\bSigma_{\k+1,d}^{-1/2},\cdots,{\bSigma}_{K,d}^{-1/2},\Q_d\}.
\end{align*}
Then by the propositions in Section \ref{appsubsec:tensor-prop} of the Online Appendix, we obtain that,
\begin{align}
	(\bscZ^o_{1,d},\cdots, \bscZ^o_{n_d-1,d})\sim \TN(\0,\I_{m_1}\circ\cdots\circ \I_{m_{\k-1}}\circ\bSigma_{\k,d}\circ \I_{m_{\k+1}}\cdots\circ \I_{m_K}\circ \I_{n_d-1}),
	\label{eqn:trans-stand-normal}
\end{align} 
which enables us to pool all i.i.d. mode-$\k$ fibers of $\{(\bscZ^o_{1,d},\cdots, \bscZ^o_{n_d-1,d}):d=1,2\}$ for the subsequent inference. 

However, the nuisances are usually unknown in practice, and we thus turn to their consistent estimates. Specifically, for $k\in[K]\bs\k,d=1,2$, {let $\hat{\bSigma}_{k,d}$ be some consistent estimate of ${\bSigma}_{k,d}$,} and we can transform the samples by
\begin{align*}
	\bscZ_{d}=\bscY_{d}\times\left\{\hat{\bSigma}_{1,d}^{-1/2},\cdots,\hat{\bSigma}_{\k-1,d}^{-1/2},\I_{\mk},\hat{\bSigma}_{\k+1,d}^{-1/2}\cdots,\hat{\bSigma}_{K,d}^{-1/2},\Q_d\right\}.
\end{align*}
Then all mode-$\k$ fibers of $(\bscZ_{1,d},\cdots,\bscZ_{n_d-1,d})$ can be seen as nearly i.i.d. observations if the covariances are appropriately estimated, and hence can be pooled to assist the inference. 
We summarize the above transformations in Algorithm \ref{alg:sample-trans}.

\begin{remark}\label{rem1}
	First, the original sample size $n_d$ should be larger than one for a proper sample transformation according to Equation \eqref{eqn:trans-stand-normal}.
	Second, though the covariance of each mode is only identifiable up to a constant, such non-identifiability will not affect the test due to the standardization step in the following statistics construction; see more details in Section \ref{appsec:stat-construct} of the Online Appendix. 
	{Third, for each $k\neq \k$, all mode-$k$ fibers can be pooled to estimate the corresponding nuisance covariance, and therefore consistent nuisance estimations described in the first step of Algorithm \ref{alg:sample-trans} are easily attainable; see examples in Remark \ref{rem3} of Section \ref{subsec:asymp-stat}.
	One can also directly pool correlated fibers in the original samples $\bscY_d$ to estimate the dependence structure of interest, it nevertheless introduces a non-negligible bias in the following variance estimations in quantifying test heterogeneities.
	As shown later in Section \ref{sec:simu}, the proposed data-driven algorithm indeed well emulates the oracle case with known nuisances and it outperforms the competing methods without such sample transformation.}
\end{remark}

\begin{algorithm}[t!]	
	\caption{\textbf{(Step 1)} Sample transformation.}
	\label{alg:sample-trans}
	\begin{description}
		\item[Input:] Original samples $\{(\bscY_{1,d},\dots,\bscY_{n_d,d}):d=1,2\}$.
		\begin{enumerate}[label*=\arabic*.]
			\item{{\bf Nuisance estimation:} Obtain $\{\hat{\bSigma}_{k,d}:k\in[K]\bs \k, d=1,2\}$ that satisfy Condition \ref{A2} in Section \ref{subsec:asymp-stat}.}	
			\item{\bf Orthogonalization:} Generate orthogonal matrices $\{\Q_d\in\bR^{n_d\times n_d}:d=1,2\}$, with the last row equal to $({1}/{\sqrt{n_d}},\ldots, {1}/{\sqrt{n_d}})$.

			\item {\bf Decorrelation and Centralization:} For $l \in [n_d], d=1,2$, calculate
			\begin{align*}
				\bscZ_{l,d}=\bscY_{l,d}\times\left\{\hat{\bSigma}_{1,d}^{-1/2},\cdots,\hat{\bSigma}_{\k-1,d}^{-1/2},\I_{\mk},\hat{\bSigma}_{\k+1,d}^{-1/2},\cdots,\hat{\bSigma}_{K,d}^{-1/2},\Q_d\right\}.
			\end{align*}
		\end{enumerate}
		\item[Output:] Transformed samples  $\{(\bscZ_{1,d},\ldots,\bscZ_{n_d-1,d}): d=1,2\}$.
	\end{description}
\end{algorithm}

\subsection{Step 2: Statistic Pairs Construction}\label{subsec:stat-pair-construct}
Based on the transformed samples $\{(\bscZ_{1,d},\cdots,\bscZ_{n_d-1,d}): d=1,2\}$, we next construct the statistic pairs $\{(T_{i,j},U_{i,j}):(i,j)\in\cH\}$ in this section. 

Recall that we focus on the comparison of the dependence structures $\{\P_{\k,d},d=1,2\}$ for a specific mode $\k$. We will consider two scenarios in turn, first {$\P_{\k,d} =\R_{\k,d}^{\scriptscriptstyle (1)}$ (hereinafter the correlation scenario) and then $\P_{\k,d} =\R_{\k,d}^{\scriptscriptstyle (2)}$ (hereinafter the partial correlation scenario)}. 


Under both scenarios, we will construct a set of primary $t$-statistics as well as an auxiliary sequence that captures the sparsity information. 
Specifically, based on the transformed samples output by Algorithm \ref{alg:sample-trans}, we estimate the correlation and partial correlation coefficients by $\hat{\rho}_{i,j,d}$'s (i.e., Equation \eqref{eqn:rho_hat_corr} in Algorithm \ref{alg:stat-corr} and Equation \eqref{eqn:rho_hat} in Algorithm \ref{alg:stat-pcorr}), whose variances will be further approximated by $\hat{\nu}_{i,j,d}$'s to handle the heterogeneity (i.e., Equation \eqref{eqn:nu_hat_corr} in Algorithm \ref{alg:stat-corr} and Equation \eqref{eqn:var_rho_hat} in Algorithm \ref{alg:stat-pcorr}). Then based on these estimates, we construct a pair of statistics $(T_{i,j}, U_{i,j})$ for each single hypothesis:
\begin{align}
	{T_{i,j}}=\frac{\hat{\rho}_{i,j,1}-\hat{\rho}_{i,j,2}}{\big(\hat{\nu}_{i,j,1}+\hat{\nu}_{i,j,2}\big)^{1/2}},\; {U}_{i,j}=\frac{\hat{\rho}_{i,j,1}+\hat{\kappa}_{i,j}\hat{\rho}_{i,j,2}}{\big(\hat{\nu}_{i,j,1}+\hat{\kappa}_{i,j}^2\hat{\nu}_{i,j,2}\big)^{1/2}},
	(i,j)\in\cH,\label{eqn:stat-pairs}
\end{align}
where $\hat{\kappa}_{i,j}=\hat{\nu}_{i,j,1}/\hat{\nu}_{i,j,2}$. 
The detailed constructions are provided in Algorithm \ref{alg:stat-corr} (the correlation scenario) and Algorithm \ref{alg:stat-pcorr} (the partial correlation scenario); more insights on the estimations in each of these two algorithms are collected in Section \ref{appsec:stat-construct} of the Online Appendix.
It is important to note that, 
the primary sequence $\{T_{i,j}\}$ in \eqref{eqn:stat-pairs} collects $t$-statistics that quantify the signal strengths of the multiple testing problem \eqref{eqn:hypothesis}, while
the auxiliary sequence $\{U_{i,j}\}$ in \eqref{eqn:stat-pairs} reflects the sparsity heterogeneity of the testing units and hence can be employed to adjust and rerank the $p$-values obtained through the primary statistics in order to improve the testing efficiency; this will be explained further in Section \ref{subsec:sera-alg}. In addition, the construction in \eqref{eqn:stat-pairs} guarantees the asymptotic independence between the two sequences (as will be shown in Theorem \ref{thm:asymp-norm_stat} below), which is essential for the subsequent testing validity analysis.


\begin{algorithm}[h!]	
	\caption{\textbf{(Step 2 for correlation  scenario)} Statistic pairs construction.}
	\label{alg:stat-corr}
	\begin{description}
		\item[Input:] Transformed samples  $\{(\bscZ_{1,d},\cdots,\bscZ_{n_d-1,d}): d=1,2\}$ from Algorithm \ref{alg:sample-trans}.
		\begin{enumerate}
			\item{\bf Estimate correlation:} Estimate the mode-$\k$ correlation matrices by
			\begin{align}
				\hat{\rho}_{i,j,d}=\frac{\hat{\sigma}_{i,j,d}}{(\hat{\sigma}_{i,i,d}\hat{\sigma}_{j,j,d})^{1/2}}, 1\leq i\leq j\leq \mk, d=1,2,
				\label{eqn:rho_hat_corr}
			\end{align}
			where
			\begin{align*}
				(\hat{\sigma}_{i,j,d})=\frac{\mk}{(n_d-1)m}\sum_{l=1}^{n_d-1}\sum_{i_k=1,k\in[K]\bs\k}^{m_k}\bscZ_{\{i_k\}_{k\in[K]\bs\k},l,d}{\bscZ}\trans_{\{i_k\}_{k\in[K]\bs\k},l,d}.
			\end{align*}

			\item{\bf Quantify heterogeneity:} Estimate the variance of $\hat{\rho}_{i,j,d}$ by
			\begin{align}
				\hat{\nu}_{i,j,d}=\frac{{\sum_{l=1}^{n_d-1}\sum_{i_k=1,k\in[K]\bs\k}^{m_k}}\left(\scZ_{i,\{i_k\}_{k\in[K]\bs\k},l,d}\scZ_{j,\{i_k\}_{k\in[K]\bs\k},l,d}-\hat{\sigma}_{i,j,d}\right)^2}{\{(n_d-1)m/\mk\}^2\hat{\sigma}_{i,i,d}\hat{\sigma}_{j,j,d}}.
				\label{eqn:nu_hat_corr}
			\end{align}

		\end{enumerate}
		
		\item[Output:] Collection of statistic pairs $\{(T_{i,j},U_{i,j}):(i,j)\in\cH\}$ via Equation \eqref{eqn:stat-pairs}.
	\end{description}
\end{algorithm}

\begin{algorithm}[t!]	
	\caption{\textbf{(Step 2 for partial correlation  scenario)} Statistic pairs construction.}
	\label{alg:stat-pcorr}	
	\begin{description}
		\item[Input:] Transformed samples  $\{(\bscZ_{1,d},\cdots,\bscZ_{n_d-1,d}): d=1,2\}$ from Algorithm \ref{alg:sample-trans}.
		
		\begin{enumerate}[label*=\arabic*.]
			\item{{\bf Estimate partial covariance:} 
			Obtain regression coefficient estimates $\{\hat{\betaa}_{i,d}\}$ that satisfy Condition \ref{A3'} in Section \ref{subsec:asymp-stat}, and calculate residuals $\hat{\xi}_{i,\{i_k\}_{k\in[K]\bs\k},l,d}=\scZ_{i,\{i_k\}_{k\in[K]\bs\k},l,d}-\bscZ_{-i,\{i_k\}_{k\in[K]\bs\k},l,d}\trans\hat{\betaa}_{i,d}$, $i\in[\mk],d=1,2$.}
			
			Calculate sample covariance of  residuals:
			\begin{align*}
				\tilde{r}_{i,j,d}=\frac{\mk}{(n_d-1)m}&{\sum_{l=1}^{n_d-1}\sum_{{i_k=1},k\in[K]\bs\k}^{m_k}}\left(\hat{\xi}_{i,\{i_k\}_{k\in[K]\bs\k},l,d}\hat{\xi}_{j,\{i_k\}_{k\in[K]\bs\k},l,d}\right).
			\end{align*}
			\item{\bf Debiasing:} Debias $\tilde{r}_{i,j,d}$ by
				\begin{align*}
					\begin{split}
						\hat{r}_{i,j,d}&=-(\tilde{r}_{i,j,d}+\tilde{r}_{i,i,d}\hat{\eta}_{i,j,d}+\tilde{r}_{j,j,d}\hat{\eta}_{j-1,i,d}),\; (i,j)\in\cH, \\
						\hat{r}_{i,i,d}&=\tilde{r}_{i,i,d},\; i \in [\mk].
					\end{split}
				\end{align*} 
			
			\item{\bf Estimate partial correlation:} 
			Calculate
			\begin{align}
				\hat{\rho}_{i,j,d}=\frac{\hat{r}_{i,j,d}}{(\hat{r}_{i,i,d}\hat{r}_{j,j,d})^{1/2}},\; (i,j)\in\cH,d=1,2.
				\label{eqn:rho_hat}
			\end{align}
			
			\item{\bf Quantify heterogeneity:} Estimate the variance of $\hat{\rho}_{i,j,d}$ by
			\begin{align}
				\hat{\nu}_{i,j,d}=(1+\hat{\eta}^2_{i,j,d}\hat{r}_{i,i,d}/\hat{r}_{j,j,d})/\{(n_d-1)m/\mk\}. 
				\label{eqn:var_rho_hat}
			\end{align}
	
		\end{enumerate}
		
		\item[Output:] Collection of statistic pairs $\{(T_{i,j},U_{i,j}):(i,j)\in\cH\}$ via Equation \eqref{eqn:stat-pairs}.
	\end{description}
\end{algorithm}

\begin{remark}
	A few remarks are collected. 
	The estimation step \eqref{eqn:rho_hat_corr} in Algorithm \ref{alg:stat-corr} is flexible and it is 
	not restricted to the sample correlation approach provided here. Other consistent correlation estimations that satisfy certain convergence rate can be employed as well \citep[e.g.,][]{cai2016large}.
	In comparison, for partial correlation case in Algorithm \ref{alg:stat-pcorr}, due to the bias introduced by node-wise regression, the {two-sample} inference of the partial correlation matrices is much more involved. We extend the ideas proposed in \cite{xia2015testing, xia2019matrix} to estimate the partial correlations. Though there exist some other partial correlation inference methods \citep[e.g.,][]{chen2019graph,lyu2019tensor}, they all target one-sample inference and cannot directly deal with two-sample cases. Nevertheless, we extend their work to the two-sample scenarios in Section \ref{sec:simu} (without theoretical guarantee) and our method presents superior performance in terms of both FDR control and power compared to theirs. The detailed description of such extension is provided in Section \ref{appsec:add-simu} of the Online Appendix. {Finally, Condition \ref{A3'} for the node-wise regression estimates can be easily satisfied and will be discussed further in Remark \ref{rem3}.}
\end{remark}

\subsection{Step 3: A Power Enhanced Procedure SERA}\label{subsec:sera-alg}
We propose in this section the procedure SERA that further improves the power of the test. Note that, SERA can be generally applied to any two-sample large-scale inference problems with sparsity structure and is not restricted to our tensor setting. Hence it is of independent interest and will be discussed further in Section \ref{subsec:sera-indep}.
We now describe the main idea of SERA and the details will be summarized in Algorithm \ref{alg:sera}. 
To be specific, we first introduce and estimate a sparsity level, and then obtain a sequence of weighted $p$-values. Finally, we approximate and control the FDP. 


Recall that, the auxiliary sequence reflects the sparsity heterogeneity among the tests. 
Hence, we incorporate such sequence and
define a \emph{posterior local sparsity level} providing the auxiliary variable $U_{i,j}$ by
\begin{align*}
	\pi(U_{i,j})=\Pr(\theta_{i,j}=1|U_{i,j}),
\end{align*}
where $\theta_{i,j}=\id(\rho_{i,j,1}\neq \rho_{i,j,2})$ for each hypothesis and $\id(\cdot)$ is an indicator function.
This sparsity level reflects the chance of signal occurrence and the goal of SERA is to use such quantity to adjust the significance of the tests.
Since $\pi(U_{i,j})$ is unknown, the first step of SERA estimates $\pi(U_{i,j})$ by $\hat{\pi}^{\tau}(U_{i,j})$ using a kernel-based approach following similar ideas in \cite{cai2021laws} and \cite{ma2022napa}, where $\tau$ is a pre-specified screening parameter; see Equation \eqref{eqn:pi-tauhat} in Algorithm \ref{alg:sera}. 

Next, we calculate the $p$-values by the asymptotic normality result of the primary sequence $\{T_{i,j}\}$ (Theorem \ref{thm:asymp-norm_stat} in Section \ref{subsec:asymp-stat}), and construct a set of locally adaptive weights based on the auxiliary sequence $\{U_{i,j}\}$. We employ the weighting scheme proposed in \cite{cai2021laws, ma2022napa}, namely, the weights $\hat{w}(U_{i,j})={\hat{\pi}^{\tau}(U_{i,j})}/\{1-\hat{\pi}^{\tau}(U_{i,j})\}$ are placed on each of the $p$-values and the adjusted $p$-values are obtained by $\{p^{\hat w}_{i,j}=p_{i,j}/\hat{w}(U_{i,j}): (i,j)\in\cH\}$. 
Intuitively, a larger $\pi(U_{i,j})$ indicates a higher chance of signal occurrence and leads to a smaller weighted $p$-value, and hence yields a higher rejection possibility for the hypothesis with index $(i,j)$. 
Such weighting scheme provides a better ranking of the tests by incorporating the sparsity information from the auxiliary sequence.
It is {worth} noting that, the weighting approach in Step 2(a) of  Algorithm \ref{alg:sera} can be flexible. Besides the approach provided here, one can employ other methods as well \citep[e.g.,][]{LiBar19,liang2023locally}.

Finally, we select a cutoff for the weighted $p$-values so that the estimated FDP does not exceed a pre-specified significance level $\alpha\in(0,1)$. Denote by $\delta^{\scriptscriptstyle{\mathtt{T}\texttt{-}\mathtt{SERA}}}_{i,j}( t) = \id (p^{\hat{w}}_{i,j} \leq t )$ the decision rule of T-SERA for the $(i,j)$-th hypothesis with threshold $t$; $\delta^{\scriptscriptstyle{\mathtt{T}\texttt{-}\mathtt{SERA}}}_{i,j}( t) = 1$ if we reject the null and $\delta^{\scriptscriptstyle{\mathtt{T}\texttt{-}\mathtt{SERA}}}_{i,j}( t) = 0$ otherwise. Then,
\begin{equation}
	\mathrm{FDP}\left\{\bdelta^{\scriptscriptstyle{\mathtt{T}\texttt{-}\mathtt{SERA}}}(t)\right\} = \frac{\sum_{(i,j)\in\cH}\left[\{1-\theta_{i,j}\} \delta^{\scriptscriptstyle{\mathtt{T}\texttt{-}\mathtt{SERA}}}_{i,j}( t)\right]}{\max\left\{\sum_{(i,j)\in\cH} \delta^{\scriptscriptstyle{\mathtt{T}\texttt{-}\mathtt{SERA}}}_{i,j}( t), 1\right\}}, \quad
	\label{eqn:fdp}
	\text{FDR}^{\scriptscriptstyle{\mathtt{T}\texttt{-}\mathtt{SERA}}} = \ep\left[ \mathrm{FDP}\left\{\bdelta^{\scriptscriptstyle{\mathtt{T}\texttt{-}\mathtt{SERA}}}(t)\right\} \right],
\end{equation}
where $\bdelta^{\scriptscriptstyle{\mathtt{T}\texttt{-}\mathtt{SERA}}}(t) = \{\delta^{ \scriptscriptstyle{\mathtt{T}\texttt{-}\mathtt{SERA}}}_{i,j}(t): (i,j)\in\cH\}$ collects all decision rules under the threshold $t$. 
To estimate and control FDP, note that the denominator of FDP is known and the numerator can be estimated by approximating its expectation. {More precisely, with known $\pi(U_{i,j})$'s and a given threshold $t$, the expected number of false rejections equals to $\ep \left[\sum_{(i,j)\in\cH}\Pr \left\{\theta_{i,j}=0,\delta^{ \scriptscriptstyle{\mathtt{T}\texttt{-}\mathtt{SERA}}}_{i,j}(t)=1 | U_{i,j} \right\} \right]\approx \ep \left\{\sum_{(i,j)\in\cH}\pi(U_{i,j})t \right\}$, where the approximation comes from the asymptotic independence between $T_{i,j}$ and $U_{i,j}$ as shown in Theorem \ref{thm:asymp-norm_stat}.} Hence, the number of false rejections, i.e., the numerator of FDP, can be estimated by $\sum_{(i,j)\in\cH}\hat{\pi}^{\tau}(U_{i,j})t$. The above steps are summarized in Algorithm \ref{alg:sera}.

\begin{algorithm}[t!]
	\caption{\textbf{(Step 3)} The power enhanced procedure SERA.}
	\label{alg:sera}
	\begin{description}
		\item[Input:]
			(a) Statistic pairs $\{(T_{i,j}, U_{i,j}):(i,j)\in\cH\}$ from Algorithm \ref{alg:stat-corr} or \ref{alg:stat-pcorr};

			\quad(b) Kernel function $\bK(\cdot)$ and bandwidth $h$;
			
			\quad(c) Screen threshold $\tau$ and significance level $\alpha$.			
		
		\begin{enumerate}[label*=\arabic*.]
			\item{\bf Posterior Sparsity Level Estimation:} \hfill
			\begin{enumerate}
				\item Calculate the $p$-values:  $p_{i,j}=2\{1-\Phi(|T_{i,j}|)\}, (i,j)\in\cH$, where $\Phi(\cdot)$ is the cumulative distribution function (CDF) of a standard normal variable;
				
				\item Determine the screen set:  $\cI(\tau)=\{(i', j')\in\cH:p_{i',j'}>\tau \}$;
				
				\item Kernel estimation:
				\begin{equation}
					\hat{\pi}^{\tau}(U_{i,j})=1-\frac{\sum_{(i',j')\in \cI(\tau)} v_{h}\left(U_{i,j},U_{i',j'}\right)}{(1-\tau) \sum_{(i',j')\in\cH} v_{h}\left(U_{i,j},U_{i',j'}\right)},
					\label{eqn:pi-tauhat}
				\end{equation}
				where \begin{math}
					v_{h}\left(U_{i,j},U_{i',j'}\right)=\frac{\bK_{h}(U_{i,j}-U_{i',j'})}{\bK_{h}(0)}, \bK_{h}(x)=1/h\bK(x/h).
				\end{math}
			
			\end{enumerate}

			\item{\bf Reranking and Thresholding:} \hfill
			\begin{enumerate}
				\item Weigh $p$-values by ${p}_{i,j}^{\hat{w}}=p_{i,j}/\hat{w}(U_{i,j})$, where $\hat{w}(U_{i,j})=\frac{\hat{\pi}^\tau(U_{i,j})}{1-\hat{\pi}^\tau(U_{i,j})},(i,j)\in\cH$; 
				
				\item\label{eqn:rank}Rank the weighted $p$-values in ascending order: ${p}^{\hat{w}}_{(1)},\cdots, {p}^{\hat{w}}_{(|\cH|)}$, where $|\cH|$ is the cardinality of $\cH$;
				
				\item Search for 
				$\hat{q}=\max\left\{q \in [|\cH|]:q^{-1}\sum_{(i,j)\in\cH}\hat{\pi}(U_{i,j}){p}^{\hat{w}}_{(q)}\leq \alpha \right\}$;
				\item Determine the decision $\delta^{\scriptscriptstyle{\mathtt{T}\texttt{-}\mathtt{SERA}}}_{i,j}(p^{\hat{w}}_{(\hat{q})})= \id (p^{\hat{w}}_{i,j} \leq p^{\hat{w}}_{(\hat{q})} )$ for each $(i,j)\in\cH$.
			\end{enumerate}
		\end{enumerate}	
		
		\item[Output:] Collection of decisions  $\bdelta^{\scriptscriptstyle{\mathtt{T}\texttt{-}\mathtt{SERA}}}(p^{\hat{w}}_{(\hat{q})})= \{\delta^{\scriptscriptstyle{\mathtt{T}\texttt{-}\mathtt{SERA}}}_{i,j}(p^{\hat{w}}_{(\hat{q})}) : (i,j)\in\cH\}$.
		
	\end{description}
\end{algorithm}

\section{Theoretical Properties}\label{sec:theory}
To better interpret the procedures in the previous section, we study in Section \ref{subsec:asymp-stat} the asymptotic properties of the statistic pairs \eqref{eqn:stat-pairs} including the asymptotic normality and independence results. Next, the error rates control of the proposed T-SERA will be explored {in Section \ref{subsec:fdr-control-sera}} and it starts with the estimation consistency result of the posterior local sparsity level {in Section \ref{subsec:pitau-consistency}.} 

We begin with some notation. For two sequences of real numbers $\{a_{n}\}$ and $\{b_{n}\}$: write $a_{n} = O(b_{n})$ if there exists a constant $C$ such that $|a_{n}| \leq C|b_{n}|$ for any sufficiently large $n$; 
write $a_{n}\asymp b_{n}$ if there exists constants $0<c<C$ such that $c|b_{n}| \leq |a_{n}| \leq C|b_{n}|$ for any sufficiently large $n$; and write $a_{n} = o(b_{n})$ if $\lim_{n\rightarrow\infty}a_{n}/b_{n} = 0$. 
For a matrix $\A$, denote by $\lambda_g(\A)$ the $g$-th largest eigenvalue of $\A$.
Let $n=n_1+n_2$ and $\tilde{m}=\max(\{m_k\}_{k\neq \k})$. 
For $l\in [n_d-1]$, denote by $\Cov^{-1}(\scZ_{\{i_k\}_{k\in[K]\bs\k},l,d})=(\tilde{\omega}^l_{i,j,d})$ and $s_0=\max_{i,l,d}\sum_{j=1}^{\mk}\max\{\id(\tilde{\omega}^l_{i,j,d}\neq 0),\id(\omega_{i,j,d}\neq 0)\}$;
{let $\betaa_{i,\{i_k\}_{k\in[K]\bs\k},d}$ be the true coefficient by regressing $\scZ_{i,\{i_k\}_{k\in[K]\bs\k},l,d}$ on $\bscZ_{-i,\{i_k\}_{k\in[K]\bs\k},l,d}$}.

\subsection{Asymptotic Normality and Independence of Statistic Pairs}\label{subsec:asymp-stat}
This section first collects some regularity conditions, and then establishes the asymptotic normalities and asymptotic independence of $T_{i,j}$ and $U_{i,j}$ defined in Equation \eqref{eqn:stat-pairs}.

\begin{enumerate}[label=(A\arabic*), series=A]	
	\item \label{A1} Suppose $n_1\asymp n_2, n_d\geq 2$, and for any $k\in[K]$, {$\lambda_{1}(\bSigma_{k,d})\asymp \lambda_{m_k}(\bSigma_{k,d})\asymp 1$}, $d=1,2$.
	
	\item \label{A2}
	Assume that uniformly in $k\in[K]\bs\k$, for some constant $C_1>5$, there exist some constants $C_0,c_{k,d}>0$ such that
	\begin{align*}
		\Pr\left\{\left\|\hat{\bSigma}_{k,d}-c_{k,d}\bSigma_{k,d}\right\|_{\infty}\geq C_0\left\{\log m_k/(n_dm/m_k)\right\}^{1/2}\right\}=O(\mk^{-C_1}).
	\end{align*}
	
	\item\label{A3}
	Suppose $\log\mk=o\{(nm/\mk)^{1/5}\}$ and $\tilde{m}^3\log \tilde{m}\log\mk\log^2\max(\tilde{m},\mk,n)=o(n\mk)$.
	
	\item[\mylabel{A3'}{(A3')}]
	Suppose $\log\mk=o\{(nm/\mk)^{1/5}\}$, 
		$s_0^2\tilde{m}^3\log \tilde{m}\log^3\max(\tilde{m},\mk,n)=o(n\mk),$ and
		$\{s_0\tilde{m}^3\log\tilde{m}\log\max(\tilde{m},\mk,n)\}^2m\log \mk =o(n\mk^3)$.
	Suppose $\{\hat{\betaa}_{i,d}\}$ satisfy that
	\begin{align*}
		\Pr\left\{\max_{i \in [\mk], \{i_k \in [m_k]\}_{k\in[K]\backslash\k}}|\hat{\betaa}_{i,d}-{\betaa_{i,\{i_k\}_{k\in[K]\bs\k},d}}|_{1}\geq a_{m1}\right\}=O(\mk^{-C_1}),\\
		\Pr\left\{\max_{i \in [\mk], \{i_k \in [m_k]\}_{k\in[K]\backslash\k}}|\hat{\betaa}_{i,d}-\betaa_{i,\{i_k\}_{k\in[K]\bs\k},d}|_{2}\geq a_{m2}\right\}=O(\mk^{-C_1}),
	\end{align*}	
	where $a_{m1}=o\left(\{\log \max(\mk,\tilde{m},n)\}\inv\right), a_{m2}=o(\left\{(nm/\mk)\log \mk\right\}^{-1/4})$.
\end{enumerate}
	 
\begin{remark}\label{rem3}
	Condition \ref{A1} states some covariance regularity conditions that are commonly assumed in the high-dimensional inference literatures on vector/matrix/tensor covariance/precision matrices \citep[e.g.,][]{bickel2008regularized, leng2012sparse, chen2019graph, lyu2019tensor}. 
	{We assume $n_d\geq 2,d=1,2$ for a proper sample transformation as explained in Remark \ref{rem1}. It is not required if the original samples are centered \citep[as assumed in, e.g.,][]{xia2017hypothesis,xia2019matrix,lyu2019tensor}. }
	Condition \ref{A2} poses some rate requirements on the nuisance covariance estimates, which can be easily satisfied by, e.g., sample covariance estimator, the banded estimator \citep{bickel2008regularized} or adaptive thresholding estimator \citep{cai2011adaptive}, under some mild conditions. Note that, the test statistics in \eqref{eqn:stat-pairs} are not affected by any deterministic constants $\{c_{k,d}:k\in[K]\bs\k,d=1,2\}$ as discussed in Remark \ref{rem1}.
	{Alternatively, one can also employ any precision estimates satisfying $\Pr\{\|\hat{\bOmega}_{k,d}-c_{k,d}^{-1}\bOmega_{k,d}\|_{2}\geq Cm_k\{\log m_k /(n_d m/m_k)\}^{1 / 2}\}=O(\mk^{-C_1})$ and the sample transformation in Algorithm \ref{alg:sample-trans} can be performed by $\bscZ_{d}=\bscY_{d}\times\{\hat{\bOmega}_{1,d}^{1/2},\cdots,\hat{\bOmega}_{\k-1,d}^{1/2},\I_{\mk},\hat{\bOmega}_{\k+1,d}^{1/2}\cdots,\hat{\bOmega}_{K,d}^{1/2},\Q_d\}$ instead.} 
	Conditions \ref{A3} and \ref{A3'} {assume} some relationships between the sample size and the tensor dimensionalities and are similar to or milder than those in the matrix scenarios \citep{chen2021testing,xia2017hypothesis, xia2019matrix}. 
	Both conditions can be further relaxed if some additional covariance sparsity assumptions are imposed. 
	Condition \ref{A3'} also regularizes the regression coefficient estimates that can be obtained via Dantzig selector, Lasso, etc., under some mild conditions.
\end{remark}

Let $\cH_{0}=\left\{(i,j)\in\cH:\theta_{i,j}=0\right\}$ and $G(t)=2\{1-\Phi(t)\}$.
{Denote by $\hat{\rho}^o_{i,j,d}$ the oracle counterpart of $\hat{\rho}_{i,j,d}$ 
that is derived from the oracle transformed samples $\{\bscZ^o_{l,d}\}_{l=1}^{n_d-1}$ in  \eqref{eqn:trans-stand-normal}, and denote by $\nu_{i,j,d}$ the theoretical variance of $\hat{\rho}^o_{i,j,d}$; the detailed expressions are provided in Section \ref{appsec:stat-construct} of the Online Appendix.}
Then, we have the following results.
\begin{theorem}\label{thm:asymp-norm_stat}
    Suppose Conditions \ref{A1}, \ref{A2} and \ref{A3} hold if $\P_{\k,d}=\R_{\k,d}^{\scriptscriptstyle (1)}$, or Conditions \ref{A1}, \ref{A2} and \ref{A3'} hold if $\P_{\k,d}=\R_{\k,d}^{\scriptscriptstyle (2)}$. Let $\rho_{i,j,1}=\rho_{i,j,2}=\rho_{i,j}$ under the null. Then,
    \begin{align}
        \begin{split}
&\Pr\left\{ \max_{(i,j)\in\cH_0}\left|T_{i,j} - {b_{i,j}}-\frac{{\hat{\rho}^o_{i,j,1}-}\hat{\rho}^o_{i,j,2}}{({\nu}_{i,j,1}+{\nu}_{i,j,2})^{1/2}}\right|\geq e_m \right\} = O(\mk^{-C_1}),\\
&\Pr\left\{\max_{(i,j)\in\cH_0}\left|U_{i,j} - {b_{i,j}}-{\mu}_{i,j}-\frac{\hat{\rho}^o_{i,j,1}-{\rho}_{i,j,1}+{\kappa}_{i,j}(\hat{\rho}^o_{i,j,2}-{\rho}_{i,j,2})}{({\nu}_{i,j,1}+{\kappa}_{i,j}^2{\nu}_{i,j,2})^{1/2}}\right|\geq e_m\right\} = O(\mk^{-C_1}),
        \end{split}
        \label{eqn:asymp-normal}
    \end{align}
    where ${\kappa}_{i,j}={\nu}_{i,j,1}/{\nu}_{i,j,2}$,  $e_m = o\{(\log \mk)^{-1/2}\}$, ${\mu}_{i,j}=(1+e_m^2)\frac{{\rho}_{i,j,1}+{\kappa}_{i,j}{\rho}_{i,j,2}}{\left({\nu}_{i,j,1}+{\kappa}_{i,j}^2{\nu}_{i,j,2}\right)^{1/2}}$, and {$b_{i,j}=O\left\{(\log\mk)^{1/2}\rho_{i,j}\right\}$ and satisfies
    $
        \Pr\left\{\max_{{(i,j)}\in\cS}|b_{i,j}|\geq x\right\}\leq O(|\cS|\{1-\Phi(x)\}+\mk^{-1})
    $
     uniformly in $(\log \mk)^{1/4}\leq x\leq (8\log \mk)^{1/2}$ and $\cS\subset\cH$.} In addition, for some constant $C_2>2$,
    \begin{align}
        \Pr_{(i,j)\in \cH_0}\left\{|T_{i,j}{-b_{i,j}}|\geq t|U_{i,j}\right\}=\{1+o(1)\}G(t)+O(\mk^{-C_2}),
        \label{eqn:asymp-ind}
    \end{align}
    uniformly in $|t| =O\left\{(\log \mk)^{1/2}\right\}$, {$|U_{i,j}{-b_{i,j}}-{\mu}_{i,j}| =O\left\{(\log \mk)^{1/2}\right\}$} and all $(i,j)\in \cH_0$. 
    
\end{theorem}

	Theorem \ref{thm:asymp-norm_stat} states the asymptotic normality of both $T_{i,j}$ and $U_{i,j}$, and also the asymptotic independence between them. The normality results enable us to derive the null distribution of the test statistics and handle the dependence among the hypotheses. The asymptotic independence is essential to SERA, because it ensures that the null distribution of the primary sequence is not distorted by incorporating the auxiliary covariates into the inference. 
	Finally, the term $b_{i,j}$ introduces an additional bias for the two-sample setting, because $\rho_{i,j,d}$'s are not {necessarily} equal to zero under the null. It adds difficulties in extending the existing asymptotic normality results in the one-sample literatures \citep[e.g.,][]{chen2019graph,lyu2019tensor} as well as in handling the dependence structures for the subsequent FDR analysis.

\subsection{Estimation Consistency of the Posterior Sparsity Level}
\label{subsec:pitau-consistency}
Next, we show the consistency of $\hat{\pi}^\tau(U_{i,j})$ in Equation \eqref{eqn:pi-tauhat} to its theoretical counterpart 
\begin{align}
	{\pi}^{\tau}(U_{i,j})=1-\frac{\Pr\{p_{i,j}>\tau|U_{i,j}\}}{1-\tau},
	\label{eqn:pi-tau}
\end{align}
which is a conservative approximation of $\pi(U_{i,j})$. We consider such an intermediate quantity \eqref{eqn:pi-tau} because a direct estimation of $\pi(U_{i,j})$ is difficult; see the explanations in \cite{cai2021laws} and \cite{ma2022napa}. 

{Denote by $f_{i',j'}(\cdot|U_{i,j})$ the conditional density function of $U_{i',j'}$ given $U_{i,j}$ and $\H$ the Hessian matrix of $\Pr(p_{i',j'}>\tau|U_{i',j'},U_{i,j})$.} We first introduce some regularity conditions. 

\begin{enumerate}[resume* = A]

	\item \label{A5} Suppose the kernel function $\bK(x):\bR\to \bR$ is positive, unimodal and satisfies $\int_{\bR}\bK(x)=1, \int_{\bR}x\bK(x)=0$, and $\int_{\bR} x^2\bK(x)<\infty$. 

	\item \label{A4} {Let $\Lambda_{i,j}=\big\{U_{i,j}$: with probability $1-O(\mk^{-2})$, uniformly for all $(i',j')\in\cH$, $f_{i',j'}(\cdot|U_{i,j})$ is bounded, $\Pr(p_{i',j'}>\tau|U_{i',j'},U_{i,j})$ has continuous first and second partial derivatives at $(U_{i',j'},U_{i,j})\trans$, and $\lambda_g(\H)=O(1)$ for $g=1,2\big\}$. Suppose $\Pr(\Lambda_{i,j})\rightarrow 1$ uniformly for all $(i,j)\in\cH$ as $\mk\to\infty$.}

 	\item \label{A6} Suppose uniformly for all $(i,j)\in\cH$, as $\mk\to\infty$, with probability tending to $1$,
	\begin{align*}
		&\Var\left(\sum_{(i',j')\in\cH}\left[\bK_{h}(U_{i,j}-U_{i',j'})\id\left(p_{i,j}>\tau\right) \right]|U_{i,j}\right)\\
		=\; &O\left(\sum_{(i',j')\in\cH}\Var\left[ \bK_h(U_{i,j}-U_{i',j'})\id\left(p_{i,j}>\tau\right) |U_{i,j}\right] \right),
	 \end{align*} 
	and 
	\begin{align*}
		&\Var\left(\sum_{(i',j')\in\cH}\bK_{h}(U_{i,j}-U_{i',j'}) |U_{i,j}\right) 
		= O\left(\sum_{(i',j')\in\cH}\Var\left[ \bK_h(U_{i,j}-U_{i',j'}) |U_{i,j}\right] \right).
	 \end{align*}
\end{enumerate}
\noindent
Condition \ref{A5} holds naturally for the commonly used kernels, and Condition \ref{A4} regulates the first and second derivatives of the conditional CDFs and is mild too.
Condition \ref{A6} can be easily satisfied by many common dependence structures as shown in \cite{ma2022napa}. 

\begin{theorem}\label{thm:piu-consist}
    Suppose Conditions \ref{A5} to \ref{A6} hold. If $h\to 0$ and $\mk^2h\to \infty$ as $\mk\to\infty$, 
    \begin{align*}
    \hat{\pi}^{\tau}(U_{i,j})\to \pi^{\tau}(U_{i,j}) , 
    \end{align*}
in probability uniformly for all $(i,j)\in\cH$.
\end{theorem}
	 The consistency results in Theorem \ref{thm:piu-consist} leads to a valid data-driven procedure T-SERA as shown in the following section. 

\subsection{Asymptotic Error Rate Control of T-SERA}
 \label{subsec:fdr-control-sera}

Define $\cH_{\zeta}=\left\{(i,j)\in\cH: \left|\rho_{i,j,d}\right| \geq{(\log \mk)^{-2-\zeta}}, \ d=1\ \mbox{or}\ 2\right\}$.
For each $i\in [\mk]$, define ${\Gamma}_{i}(\gamma)=\left\{j: j\in [\mk],j\neq i,\left|\rho_{i,j,d}\right| \geq(\log \mk)^{-2-\gamma}, \ d=1\ \mbox{or}\ 2\right\}$.
Let $\cH_{\upsilon}=\{(i,j)\in\cH: \\ \frac{\rho_{i,j,1}-\rho_{i,j,2}}{(\nu_{i,j,1}+\nu_{i,j,2})^{1/2}}\geq (\log \mk)^{1/2+\upsilon}\}$.  
This section first collects some regularity conditions and then shows the error rate control of SERA.
\begin{enumerate}[resume* = A]
\item \label{A7}
Suppose there exists some constant $\zeta>0$ such that {$\sum_{(i,j)\in\cH_{\zeta}}\id\{\theta_{i,j}=0\}={o_{\mathsf{p}}}(\mk^{\beta})$} for any sufficiently small constant $\beta>0$. 

\item \label{A8}
Suppose $\log\mk=o\{(nm/\mk)^{1/8}\}$, and there exists {some $0<\gamma<\zeta$} such that $\max_{i\in [\mk]}\left|\Gamma_{i}(\gamma)\right| \asymp 1$, where $\zeta$ is defined in \ref{A7}. 

\item \label{A9} 
Suppose, with probability tending to 1, $\pi^{\tau}(U_{i,j}) \in[\varrho, 1-\varrho]$ for some sufficiently small constant $\varrho>0$, and has bounded first derivative with respect to $U_{i,j}$. 
Suppose that 
${\Var_{\theta_{i,j},(i,j)\in\cH}\sum_{(i,j)\in\cH} \left(\ep_{U_{i,j}}[w(U_{i,j})|\{\theta_{i,j},(i,j)\in\cH\}]\id\{\theta_{i,j}=0\}\right)=o\left(\mk^{4}\right)}$ where $w(U_{i,j})=\pi^{\tau}(U_{i,j})/\{1-\pi^{\tau}(U_{i,j})\}$.
	
\item \label{A10} Suppose there exists some $\upsilon, \varepsilon>0$ such that $\left|\cH_{\upsilon}\right| \geq [1 /\{(8c_\pi)^{1 / 2} \alpha\}+\varepsilon ] (\log\log \mk)^{1 / 2}$, where $c_\pi\approx3.14$ is a math constant. 
\end{enumerate}
\noindent
Condition \ref{A7} assumes that not too many mode-$\k$ tensor elements have strong and exactly equal correlations/partial correlations.
Condition \ref{A8} indicates that most of the test statistics are weakly correlated with each other and
Condition \ref{A9} requires that the latent variables $\theta_{i,j}$'s are not perfectly correlated.
Finally, Condition \ref{A10} is assumed to avoid an overly conservative FDR; it requires a few hypotheses to have the standardized correlation/partial correlation difference exceeding $(\log \mk)^{1/2+\upsilon}$. 
The conditions are all mild and they are similarly assumed in many high-dimensional multiple testing literatures \citep[e.g.,][]{liu2013gaussian,xia2015testing, xia2019gap, cai2021laws,ma2022napa}. 

\begin{theorem}\label{thm:fdr_control_dd}
	Suppose Conditions \ref{A1} to \ref{A10} hold. Then, for any $\epsilon > 0$, 
	\[
		\limsup_{\mk\to\infty} \operatorname{FDR}^{\scriptscriptstyle{\mathtt{T}\texttt{-}\mathtt{SERA}}} \leq \alpha, \;\; \text { and } \;\; 
		\lim _{\mk\to\infty} \Pr\left\{\operatorname{FDP}\left(\boldsymbol{\delta}^{\scriptscriptstyle{\mathtt{T}\texttt{-}\mathtt{SERA}}}\right) \leq \alpha+\epsilon\right\} = 1.
	\]
\end{theorem}
The above asymptotic FDP and FDR control results of T-SERA can be extended to more general cases as shown in the next section.

\subsection{Generality of SERA}\label{subsec:sera-indep}
We emphasize that, though we mainly focus on the tensor dependence testing problem in the current article, the proposed SERA is generally applicable to many other sparse two-sample multiple testing problems as detailed below.

Suppose there are two groups of data $\Y_d=\{Y_{i,d}\}_{i=1}^m$, $d=1,2$, that follows a probability distribution $\mathcal{P}_{\boldsymbol{\beta}_d,\boldsymbol{\eta}_d}$, where $\boldsymbol{\beta}_d=\{ \beta_{i,d} : i \in [m] \}$ denotes the parameters of interest, and $\boldsymbol{\eta}_d$ collects all the nuisance parameters. Suppose we observe two sets of independent samples, $\{\Y_{l,d} \}_{l=1}^{n_d}$, where $n_d$ is the sample size for group $d$, and one wishes to carry out multiple hypothesis testing of
\begin{equation} \label{eqn:hypothesis-sera}
H_{0,i}: \; \beta_{i,1} = \beta_{i,2} \quad \mbox{versus} \quad H_{1,i}: \; \beta_{i,1} \neq \beta_{i,2}, \qquad i \in [m].
\end{equation}
Besides the tensor problem considered in this article, \eqref{eqn:hypothesis-sera} also covers a range of two-sample testing problems including detecting differential networks, identifying gene-environment interactions, etc. 
The following corollary shows that, if the primary and auxiliary statistics $\{(T_i, U_i), i\in [m]\}$ for \eqref{eqn:hypothesis-sera} are {appropriately constructed (Condition \ref{C1} below) similarly as in \eqref{eqn:stat-pairs} for the tensor setting, 
then the decision rule $\boldsymbol{\delta}^{\scriptscriptstyle\mathtt{SERA}}$ obtained by applying SERA to $\{(T_i, U_i), i\in [m]\}$ can asymptotically control the FDP and FDR as well.}

With slight abuse of notation, we let $\cH_0=\left\{i \in [m]: \beta_{i,1} = \beta_{i,2}\right\}$ for the general two-sample testing problem \eqref{eqn:hypothesis-sera} and let $\beta_{i,1}=\beta_{i,2}=\beta_{i}$ for $i\in \cH_0$.
{Let $\psi(Y_{l,i,d})$ denote the influence function of $\beta_{i,d}$ at $Y_{l,i,d}$. Let $Z_{k,i} = (n_2/n_1)\psi(Y_{l,i,1})$, for $k = l, l\in [n_1]$, and $Z_{k,i} = -\psi(Y_{l,i,2})$, for $k = n_1 + l, l\in [n_2]$.}
\begin{enumerate}[label=(C\arabic*), series=P]
\item \label{C1}
Suppose that $n_1\asymp n_2$, $\log m=o(n^{1/8})$ and  $\ep\left\{\exp\left(C_3|Z_{k, i}|/[\Var\{Z_{k,i}\}]^{1/2}\right)\right\} < \infty$ for some $C_3 > 0$. {Suppose that there exists some ${\mu_i=\left(1+o\{(\log m)^{-1}\}\right)\ep(U_i)}$ and $b_{i}=O\left\{(\log m)^{1/2}\beta_{i}\right\}$ that satisfies 
$
    \Pr\left\{\max_{i\in\cS}|b_{i}|\geq x\right\}\leq O(|\cS|(1-\Phi(x))+m^{-1})
$
uniformly in $(\log m)^{1/4}\leq x\leq (4\log m)^{1/2}$ and  $\cS\subset[m]$, such that 
\begin{align*}
&\Pr_{H_{0,i}}\left\{ \left|T_i {-b_i}-\frac{\sum_{k=1}^{n_1+n_2}Z_{k,i}}{\Var\{\sum_{k=1}^{n_1+n_2}Z_{k,i}\}^{1/2}}\right|\geq e_m \right\} = O(m^{-C_4}),\\
&\Pr_{H_{0,i}}\left\{\left|\left[U_i {-b_i}-{{\mu}_i}\right]-\frac{\sum_{k=1}^{n_1}Z_{k,i}-\vartheta_i\sum_{k=n_1+1}^{n_1+n_2}Z_{k,i}}{\Var\{\sum_{k=1}^{n_1}Z_{k,i}-\vartheta_i\sum_{k=n_1+1}^{n_1+n_2}Z_{k,i}\}^{1/2}}\right|\geq e_m\right\} = O(m^{-C_4}),
\end{align*}
}
for some constant $C_4 > 5$ and $e_m = o\{(\log m)^{-1/2}\}$, where $\vartheta_i = \left[ n_1\Var\{Z_{1,i}\} \right]/\left[ n_2\Var\{Z_{n,i}\} \right]$.

\end{enumerate}

\begin{corollary}\label{thm:fdr_control_oracle}
	Suppose Conditions \ref{C1}, {\ref{A5}, \ref{A5'} - \ref{A10'} hold.} Then, for any $\epsilon > 0$,
	\begin{align*}
	\limsup_{m\to\infty} \operatorname{FDR}^{\scriptscriptstyle\mathtt{SERA}} \leq \alpha, \;\; \text { and } \;\; 
	\lim _{m\to\infty} \Pr\left\{\operatorname{FDP}\left(\boldsymbol{\delta}^{\scriptscriptstyle\mathtt{SERA}}\right) \leq \alpha+\epsilon\right\} = 1. 
	\end{align*}
	 
\end{corollary}
We remark that Condition \ref{C1} can be easily satisfied by our own construction; see \cite{xia2019gap} for the guidance. {Conditions \ref{A5'} to \ref{A10'} are the analogs of \ref{A4} to \ref{A10} for the general two-sample problem in \eqref{eqn:hypothesis-sera} and are presented in detail in Section \ref{appsubsec:coro-cond} of the Online Appendix.} 
Therefore, the above corollary provides a general guarantee of error rate control for applying the proposed SERA to a broad range of two-sample multiple testing problems discussed above.

\section{Simulation Studies}\label{sec:simu}
In this section, we investigate the numerical performance of the proposed \texttt{T-SERA} under the correlation and partial correlation scenarios in turn. 
In both cases, we compare \texttt{T-SERA} with \texttt{T-BH} and \texttt{T-GAP}, where \texttt{T-BH} only employs the primary tensor statistics $\{T_{i,j}\}$ in \eqref{eqn:stat-pairs} in BH procedure \citep{benjamini1995controlling} and \texttt{T-GAP} incorporates the statistic pairs in \eqref{eqn:stat-pairs} in GAP procedure \citep{xia2019gap}. Note that, GAP adopts the auxiliary statistics in a discrete grouping fashion and is expected to be less efficient and computationally slower.
In the partial correlation scenario, we additionally compare \texttt{T-SERA} with two modified tests from their one-sample versions \citep{chen2019graph, lyu2019tensor} (by first constructing similar statistic pairs as in \eqref{eqn:stat-pairs} and then applying the newly proposed SERA), and the two tests are denoted by \texttt{CL-SERA} and \texttt{LX-SERA}, respectively.
The detailed descriptions of such modifications are relegated to Sections \ref{appsubsec:stat-cl} and \ref{appsubsec:stat-lx} of the Online Appendix. 
For the proposed method, we additionally compare the data-driven \texttt{T-SERA} with its oracle version where the nuisances $\{\bSigma_{k,d}:k\in [K]\setminus \k;d=1,2\}$ are known {(denoted by \texttt{T-SERA}$^{\mathtt{OR}}$)}.


\subsection{Data Generation and Implementation}\label{subsec:simu-data}
In both scenarios, we generate the third-order tensor (i.e., $K=3$) samples by
\begin{align}
	\begin{split}
		\bscY_{1,1},\dots,\bscY_{n_1,1}\in\bR^{m_1\times m_2\times m_3}&\sim \TN(\bscM_1,\bSigma_{1,1}\circ\bSigma_{2,1}\circ\bSigma_{3,1}),\\ \bscY_{1,2},\dots,\bscY_{n_2,2}\in\bR^{m_1\times m_2\times m_3}&\sim \TN(\bscM_2,\bSigma_{1,2}\circ\bSigma_{2,2}\circ\bSigma_{3,2}),
	\end{split}
	\label{eqn:simu-sample}
\end{align} 
with {$m_1=100, m_2=20, m_3=10$ and $n_1=n_2=3$, which mimic the dimensionalities and sample sizes of the datasets in Section \ref{sec:real-data}.}
Denote by $\bscM_{d}=\{\scM_{i_1,i_2,i_3,d}:i_1\in[m_1],i_2\in[m_2],i_3\in[m_3]\}$. We independently generate $\scM_{i_1,i_2,i_3,1}\sim 3\operatorname{N}(0, 1)$ and $\scM_{i_1,i_2,i_3,2} \sim 2\operatorname{N}(0, 1)$. 
{Without loss of generality, let the pre-specified mode of interest $\k=1$}.  

Again, we unify the dependence structures by {$\P_{k}=(\rho_{i,j})_{m_k\times m_k}$} for the correlation and partial correlation scenarios.
Two structural collections of {$\P_{k}$} are considered in the following generation process.
The first collection consists of three sparse structures, i.e., {\texttt{Band}: $\rho_{i, i} = 1$ for $i\in[m_k]$, $\rho_{i,j} = 0.6$ for $|i-j|=1$, $\rho_{i,j}=0.3$ for $|i-j|=2$ and $\rho_{i,j} = 0$ otherwise; \texttt{Hub}: $\rho_{i, i} = 1$ for $i\in[m_k]$, $\rho_{i,j} = \rho_{j, i} = 0.5$ for $i = 10(l - 1) + 1$, $10(l - 1) + 2 \leq j \leq 10(l - 1) + 10$ where $1 \leq l \leq m_k/10$ and $\rho_{i,j} = 0$ otherwise\footnote[1]{In the implementation, we let $\P_k \leftarrow \P_k + [|\lambda_{m_k}(\P_k)|+0.05]\I_{m_k}$ to ensure the positive definiteness.}; and \texttt{Random}: $\rho_{i,i} = 1$ for $i\in[m_k]$, $\rho_{i,j} = \rho_{j,i} = \operatorname{Uniform}(0.4, 0.8) \times \operatorname{Bernoulli}(1, \min\{0.05,10/m_k\})$\footnote[2]{The sampling of the two distributions is independent of each other.} for $1 \leq i < j \leq m_k$ and $\rho_{i,j} = 0$ otherwise\footnotemark[1].}
The second collection consists of two auto-regressive (AR) structures and two moving average (MA) structures, i.e., \texttt{AR4}: $\rho_{i,j} = 0.4^{|i-j|}$; \texttt{AR5}: $\rho_{i,j} = 0.5^{|i-j|}$; \texttt{MA3}: $\rho_{i,j} = \frac{1}{|i-j|+1}$ for $1 \leq |i-j| \leq 3$ and $\rho_{i,j} = 0$ otherwise; and \texttt{MA4}: $\rho_{i,j} = \frac{1}{|i-j|+1}$ for $1 \leq |i-j| \leq 4$ and $\rho_{i,j} = 0$ otherwise.

We then introduce the generation mechanism of mode-$k$ dependence structures for $k\in [3]$, and we start with the correlation case.
{For $k=\k=1$}, namely the mode of interest, two configurations are considered. 
For the first configuration, we select different structures from the first dependence 
collection described above for each of the two groups, for example, generate $\bSigma_{1,1}$ by \texttt{Band} and $\bSigma_{1,2}$ by \texttt{Hub}. 
For the second configuration, we first generate a correlation matrix $\R_{1}^{\scriptscriptstyle (1)} = (\rho_{i,j})$ from the first structural collection and randomly select {half of the} non-zero locations in its upper triangle.
Then, these locations are randomly divided into two sets with equal sizes, denoted by $G_1$ and $G_2$, and we let $\Delta_{d}=(\delta_{i,j}), \delta_{i,j}=\rho_{i,j}$ if $(i,j)\in G_{d}$ and $\delta_{i,j}=0$ otherwise, for $d=1,2$.
Finally, we set $\bSigma_{1,d}=\R_{1}^{\scriptscriptstyle (1)}+\Delta_{d}+(\varsigma+0.05)\I_{m_1}$ for $d=1,2$, where $\varsigma=|\min_{d=1,2}\left\{\lambda_{m_1}\left(\R_{1}^{\scriptscriptstyle (1)}+\Delta_{d}\right)\right\}|$, and the dependence structures of the two groups are respectively denoted by the superscripts $\mathtt{G_1}$ and $\mathtt{G_2}$ (e.g., \texttt{Band}$^\mathtt{G_1}$, \texttt{Band}$^\mathtt{G_2}$).
For $k=2,3$, namely the nuisance modes,  two settings from the second structural collection are considered, i.e., \texttt{AR}: generate $\bSigma_{k, 1}$ and $\bSigma_{k, 2}$ by \texttt{AR4} and \texttt{AR5}, respectively; \texttt{MA}: generate $\bSigma_{k, 1}$ and $\bSigma_{k, 2}$ by \texttt{MA3} and \texttt{MA4}, respectively. 
For partial correlation case, for $k=1$, same generation mechanism above is employed while replacing $\R_{1}^{\scriptscriptstyle (1)}$ with $\R_{1}^{\scriptscriptstyle (2)}$ and replacing $\bSigma_{1, d}$ with $\bOmega_{1, d}$; for $k=2,3$, exactly the same generation process as the correlation case is adopted.

Next, we describe the implementation details of the proposed method.
We estimate the nuisances by the sample covariance matrices for simplicity and computation efficiency. The bandwidth $h$ for estimating $\pi^{\tau}(U_{i,j})$ in \eqref{eqn:pi-tauhat} is selected by the {\texttt{hns}} function in the \texttt{R} package \texttt{ks}. 
We follow \cite{cai2021laws, ma2022napa} and choose $\tau$ in \eqref{eqn:pi-tauhat} as the BH threshold at the significance level $0.9$. 
To stabilize the estimation, we truncate $\hat{\pi}^{\tau}(U_{i,j}) = \xi$ if $\hat{\pi}^\tau(U_{i,j})<\xi$, and $\hat{\pi}^{\tau}(U_{i,j}) = 1-\xi$ if $\hat{\pi}^\tau(U_{i,j})>1-\xi$, where we set $\xi=10^{-5}$. In addition, under the partial correlation scenario, node-wise lasso is applied to obtain the regression coefficient estimates in Algorithm \ref{alg:stat-pcorr} and
the tuning parameter {is chosen following \cite{liu2013gaussian}.} 
All simulation results are based on 100 independent replications with significance level $\alpha=5\%$.


\subsection{FDR and Power Comparison}\label{subsec:simu-fdr}

Table \ref{table:corr} collects the empirical FDR and power, in percentages, of the proposed \texttt{T-SERA} as well as the competing methods \texttt{T-BH} and \texttt{T-GAP} under the correlation scenario. 
It can be seen that the proposed \texttt{T-SERA} successfully controls FDR in all data generation settings. 
Moreover, the data-driven results are close to those of the oracle procedure \texttt{T-SERA}$^{\mathtt{OR}}$; they both show some conservativeness which may partly attribute to the conservative approximation of $\pi(U_{i,j})$ as well as the choice of bandwidth $h$ in \eqref{eqn:pi-tauhat}.
Besides, \texttt{T-BH} often suffers from low power, especially for the second configuration where the covariance adopts \texttt{Hub} or \texttt{Random} structures.
This is because the magnitudes of entries generated by these two structures are usually very small. 
In comparison, \texttt{T-GAP} improves over \texttt{T-BH} because of the incorporation of the auxiliary sequence through the grouping and adjusting idea in \cite{xia2019gap}, while the proposed method shows additional superiority over \texttt{T-GAP} because \texttt{T-SERA} reranks the $p$-values by exploring the underlying sparsity structure in a continuous fashion rather than discrete grouping and thereby enjoys additional testing efficiency gain. 
} 

Table \ref{table:pcorr} presents the results for the partial correlation scenario. The comparisons of \texttt{T-SERA} with  \texttt{T-BH} and \texttt{T-GAP} are quite similar to those in Table \ref{table:corr}. 
For the two additional competing methods \texttt{CL-SERA} and \texttt{LX-SERA}, it can be seen from the table that, {both approaches are  less powerful than \texttt{T-SERA}; sometimes they cannot even compete with \texttt{T-BH}.}
Therefore, across all settings, the proposed method enjoys a superior performance compared to all four competing methods.

\begin{table}[t!]\small
	\setlength{\tabcolsep}{1.5pt}{
	\begin{tabular}{l|cccccccccccc} 
	\hline
	\multicolumn{1}{c|}{$\bSigma_{2,d}\& \bSigma_{3,d}$} & \multicolumn{6}{c|}{\texttt{AR}}& \multicolumn{6}{c}{\texttt{MA}}    \\ 
	\cline{2-13}
	\multicolumn{1}{c|}{$\bSigma_{1,1}$}  & \multicolumn{1}{c|}{\texttt{Band}} & \multicolumn{1}{c|}{\texttt{Hub}}    & \multicolumn{1}{c|}{\texttt{Random}} & \multicolumn{1}{c|}{\texttt{Band$^{\mathtt{G_1}}$}} & \multicolumn{1}{c|}{\texttt{Hub$^{\mathtt{G_1}}$}} & \multicolumn{1}{c|}{\texttt{Random$^{\mathtt{G_1}}$}} & \multicolumn{1}{c|}{\texttt{Band}} & \multicolumn{1}{c|}{\texttt{Hub}}    & \multicolumn{1}{c|}{\texttt{Random}} & \multicolumn{1}{c|}{\texttt{Band$^{\mathtt{G_1}}$}} & \multicolumn{1}{c|}{\texttt{Hub$^{\mathtt{G_1}}$}} & \texttt{Random}$^{\mathtt{G_1}}$  \\ 
	\cline{2-13}
	\multicolumn{1}{c|}{$\bSigma_{1,2}$}  & \multicolumn{1}{c|}{\texttt{Hub}}  & \multicolumn{1}{c|}{\texttt{Random}} & \multicolumn{1}{c|}{\texttt{Band}}   & \multicolumn{1}{c|}{\texttt{Band$^{\mathtt{G_2}}$}} & \multicolumn{1}{c|}{\texttt{Hub$^{\mathtt{G_2}}$}} & \multicolumn{1}{c|}{\texttt{Random$^{\mathtt{G_2}}$}} & \multicolumn{1}{c|}{\texttt{Hub}}  & \multicolumn{1}{c|}{\texttt{Random}} & \multicolumn{1}{c|}{\texttt{Band}}   & \multicolumn{1}{c|}{\texttt{Band$^{\mathtt{G_2}}$}} & \multicolumn{1}{c|}{\texttt{Hub$^{\mathtt{G_2}}$}} & \texttt{Random}$^{\mathtt{G_2}}$  \\ 
	\hline
	\multicolumn{13}{c}{Empirical FDR (\%)}     \\ 
	\hline
	\texttt{T-SERA}$^{\mathtt{OR}}$  & 1.63 & 3.19 & 2.27 & 2.25 & 3.64& 2.51   & 1.53 & 2.89 & 2.36 & 2.31 & 3.17& 2.81\\
	\texttt{T-SERA} & 1.53 & 3.10 & 2.09 & 1.94 & 3.31& 2.33   & 1.48 & 2.99 & 2.28 & 2.16 & 3.46& 2.94\\
	\texttt{T-BH}   & 4.58 & 4.66 & 4.28 & 4.55 & 3.67& 5.28   & 4.57 & 4.68 & 4.56 & 4.54 & 4.46& 2.94\\
	\texttt{T-GAP}  & 4.70 & 2.82 & 2.69 & 1.83 & 1.69& 1.27   & 4.68 & 2.99 & 2.99 & 1.63 & 1.30& 1.57\\ 
	\hline
	\multicolumn{13}{c}{Empirical Power (\%)}   \\ 
	\hline
	\texttt{T-SERA}$^{\mathtt{OR}}$  & 96.46& 83.46& 87.72& 78.03& 85.87  & 46.08  & 96.35& 83.50& 88.12& 77.99& 85.84  & 45.35     \\
	\texttt{T-SERA} & 96.37& 82.80& 82.60& 78.15& 84.93  & 44.85  & 96.28& 82.87& 87.71& 77.64& 85.60  & 44.70     \\
	\texttt{T-BH}   & 89.54& 58.14& 70.31& 45.57& 5.53& 1.67   & 89.29& 57.95& 70.24& 46.42& 5.16& 1.59\\
	\texttt{T-GAP}  & 94.30& 75.86& 82.11& 76.57& 76.09  & 35.08  & 94.46& 75.73& 81.84& 76.80& 77.67  & 35.04     \\
	\hline
	\end{tabular}}
	\caption{The empirical FDR and power comparison for the correlation scenario; $\alpha=5\%$.}
	\label{table:corr}
\end{table}
	
\begin{table}[t!]\small
	\setlength{\tabcolsep}{1.5pt}{
	\begin{tabular}{l|cccccccccccc}
	\hline
	\multicolumn{1}{c|}{$\bSigma_{2,d}\& \bSigma_{3,d}$} & \multicolumn{6}{c|}{\texttt{AR}}    & \multicolumn{6}{c}{\texttt{MA}}               \\ 
	\cline{2-13}
	\multicolumn{1}{c|}{$\bOmega_{1,1}$}& \multicolumn{1}{c|}{\texttt{Band}} & \multicolumn{1}{c|}{\texttt{Hub}}    & \multicolumn{1}{c|}{\texttt{Random}} & \multicolumn{1}{c|}{\texttt{Band}$^{\mathtt{G_1}}$} & \multicolumn{1}{c|}{\texttt{Hub}$^{\mathtt{G_1}}$} & \multicolumn{1}{c|}{\texttt{Random}$^{\mathtt{G_1}}$} & \multicolumn{1}{c|}{\texttt{Band}} & \multicolumn{1}{c|}{\texttt{Hub}}    & \multicolumn{1}{c|}{\texttt{Random}} & \multicolumn{1}{c|}{\texttt{Band}$^{\mathtt{G_1}}$} & \multicolumn{1}{c|}{\texttt{Hub}$^{\mathtt{G_1}}$} & \texttt{Random}$^{\mathtt{G_1}}$  \\ 
	\cline{2-13}
	\multicolumn{1}{c|}{$\bOmega_{1,2}$}& \multicolumn{1}{c|}{\texttt{Hub}}  & \multicolumn{1}{c|}{\texttt{Random}} & \multicolumn{1}{c|}{\texttt{Band}}   & \multicolumn{1}{c|}{\texttt{Band}$^{\mathtt{G_2}}$} & \multicolumn{1}{c|}{\texttt{Hub}$^{\mathtt{G_2}}$} & \multicolumn{1}{c|}{\texttt{Random}$^{\mathtt{G_2}}$} & \multicolumn{1}{c|}{\texttt{Hub}}  & \multicolumn{1}{c|}{\texttt{Random}} & \multicolumn{1}{c|}{\texttt{Band}}   & \multicolumn{1}{c|}{\texttt{Band}$^{\mathtt{G_2}}$} & \multicolumn{1}{c|}{\texttt{Hub}$^{\mathtt{G_2}}$} & \texttt{Random}$^{\mathtt{G_2}}$  \\ 
	\hline
	\multicolumn{13}{c}{Empirical FDR (\%)}          \\ 
	\hline
	\texttt{T-SERA}$^{\mathtt{OR}}$     & 0.95 & 2.16   & 1.48   & 2.07 & 3.26& 2.29   & 0.94 & 2.30   & 1.66   & 2.34 & 3.24& 1.92\\
	\texttt{T-SERA}        & 2.31 & 4.60   & 2.55   & 3.46 & 5.09& 3.08   & 2.77 & 5.24   & 2.92   & 3.94 & 5.68& 3.32\\
	\texttt{T-BH}          & 5.58 & 6.76   & 5.13   & 5.47 & 5.18& 4.02   & 5.61 & 7.00   & 5.63   & 5.37 & 7.15& 5.25\\
	\texttt{T-GAP}         & 6.02 & 4.05   & 3.28   & 2.73 & 4.67& 1.62   & 6.41 & 4.34   & 3.75   & 3.10 & 4.75& 1.50\\
	\texttt{CL-SERA}       & 0.93 & 1.33   & 0.54   & 3.71 & 4.66& 2.99   & 1.78 & 0.98   & 0.82   & 2.99 & 2.76& 1.88\\
	\texttt{LX-SERA}       & 1.42 & 2.57   & 1.36   & 1.66 & 2.84& 1.35   & 1.09 & 2.30   & 0.84   & 1.09 & 2.99& 5.94\\ 
	\hline
	\multicolumn{13}{c}{Empirical Power (\%)}        \\ 
	\hline
	\texttt{T-SERA}$^\mathtt{OR}$       & 96.56& 88.67  & 90.79  & 85.50& 95.42& 58.98  & 96.65& 88.66  & 91.27  & 85.41& 96.27& 58.06            \\
	\texttt{T-SERA}        & 96.66& 87.32  & 90.14  & 83.57& 96.04& 56.69  & 96.74& 87.24  & 90.71  & 82.79& 95.96& 56.29            \\
	\texttt{T-BH}          & 92.88& 67.82  & 76.13  & 56.14& 38.89& 4.80   & 92.93& 67.95  & 76.78  & 56.86& 38.58& 5.18\\
	\texttt{T-GAP}         & 95.16& 81.93  & 85.84  & 82.55& 90.71& 48.10  & 95.21& 81.80  & 86.04  & 82.07& 90.58& 47.53            \\
	\texttt{CL-SERA}       & 87.15& 51.65  & 62.00  & 65.37& 46.09& 27.96  & 83.20& 57.46  & 61.09  & 61.77& 56.16& 18.10            \\
	\texttt{LX-SERA}       & 80.83& 58.06  & 66.91  & 60.05& 20.56& 9.12   & 69.09& 38.83  & 53.79  & 52.42& 5.20& 1.54\\
	\hline
	\end{tabular}}
	\caption{The empirical FDR and power comparison for the partial correlation scenario; $\alpha=5\%$.}
	\label{table:pcorr}
\end{table}

\section{Real Data Analysis}\label{sec:real-data}
In this section, we investigate the performance of the proposed method on two real datasets, an international trade dataset for detecting correlation alteration of commodity types, and a climate dataset for detecting partial correlation alteration of spatial locations in US. Significance levels are set as $\alpha=5\%$ in both studies. 

\subsection{Example of Correlation Comparison}\label{subsec:real-data-corr}
The first dataset consists of monthly imports (CIF value) of {97 commodity types over 30 countries} from the year 2015 to 2022, and is available at the UN Comtrade website \url{https://comtradeplus.un.org}. Data from each year serves as an observation, which is a {$97\times 30\times 12$} tensor.  
Such trade data have been well studied in the literatures, for example, in correlation estimation \citep{hoff2011separable} and tensor extrapolation \citep{schosser2022tensor}. Following the spirit of \cite{hoff2011separable}, the scientific interest in this section is to detect the correlation alteration of the commodity mode {(i.e., $\k=1$}), before and after \emph{COVID-19 pandemic} hit in December 2019 \citep{page2021hunt}. 
After adopting the pre-processing method in \cite{leng2012sparse} and \cite{chen2019graph} that reduces the potential serial correlations among the observations,
the first group of the processed dataset consists of the {lag-one differential tensor observations from 2016 to 2019}, and the second group consists of those from 2020 to 2022. 
The complete lists of 97 commodity types and 30 countries are collected in Section \ref{appsec:data-info-trade} of the Online Appendix. 

\begin{figure}[t!]
	\centering
	\includegraphics[scale = 0.8]{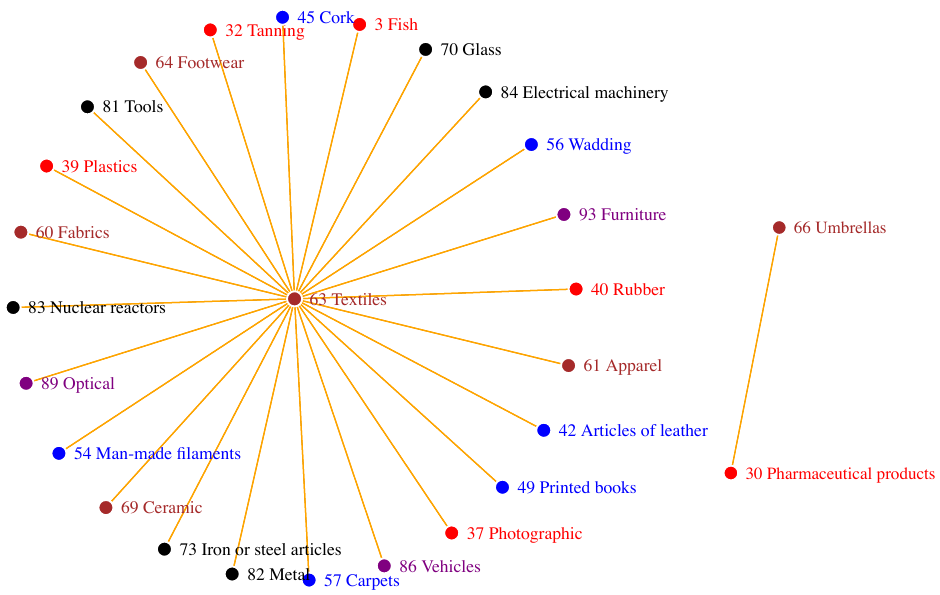}
	\caption{{Top 25} correlation alterations found by \texttt{T-SERA} in the trade example; {$\alpha=5\%$}. {Each node is a commodity type named by its index and its first phrase; see commodity descriptions in Table \ref{tab:commod} in Section \ref{appsec:data-info-trade} of the Online Appendix.}}
	\label{fig:trade}
\end{figure}

Before conducting the analysis, it is crucial to verify the separability assumption in the covariance structure. Specifically, we adopt the bootstrap method proposed in \cite{aston2017tests} and apply their test to each of the {mode-$k$ matricization, $k\in[3]$}, 
for both groups respectively. The results suggest the acceptance of the separability assumption. 
Next, we apply \texttt{T-BH}, \texttt{T-GAP} and \texttt{T-SERA} to the dataset and they yield a rejection {of 64, 92, and 98} respectively, out of a total of 4656 hypotheses. Besides, the altered correlations among commodities found by \texttt{T-SERA} covers 98.4\% of those found by \texttt{T-BH} and 97.8\% of those found by \texttt{T-GAP}. Together with our simulation studies, it suggests that the proposed \texttt{T-SERA} manages to achieve the best power. 
Note that from an economic perspective, 
imports of substitutable commodities may have a high negative correlation, while imports of complementary commodities may have a positive correlation.
However, the COVID-19 pandemic greatly impacts the global economy. For instance, countries may wish to shrink their global demands and seek for localization \citep{rajput2021shock}. Such a shift may result in a reconstruction of the commodity market and an alteration of the import correlation. 
To facilitate visualization, in Figure \ref{fig:trade} we picture the {top 25} detected alterations by \texttt{T-SERA} according to its weighted $p$-values.
We observe that {24 out of those 25} top findings are related to the commodity type ``Textiles'', which agrees with the literature conclusion that the COVID-19 pandemic impacts the international trades of the textile and fashion industry \citep{blancheton2021french,kanupriya2021covid,arania2022impact,haukkala2023fashion}. 

\subsection{Example of Partial Correlation Comparison}\label{subsec:real-data-pcorr}
\begin{figure}[t!]
	\centering
	\includegraphics{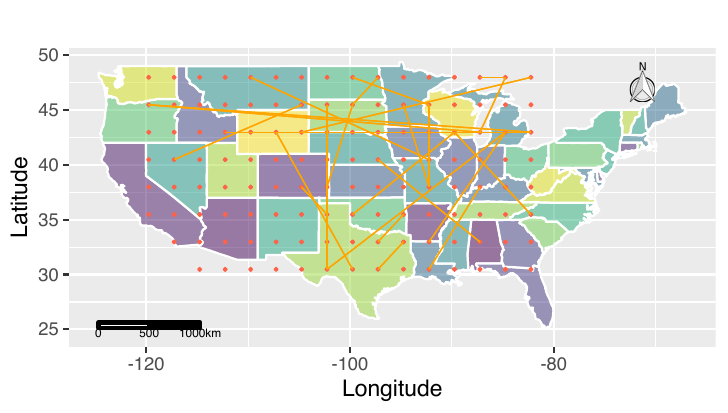}
	\caption{{Top 25} partial correlation alterations found by \texttt{T-SERA} in the climate example; {$\alpha=5\%$}; see location details in Table \ref{tab:location} in Section \ref{appsec:data-info-climate} of the Online Appendix.}
	\label{fig:cliamte}
\end{figure}

The second dataset consists of monthly measurements of 17 meteorological factors over 125 locations in US from the year 1990 to 1996, and is available at the USC Melady Lab website \url{https://viterbi-web.usc.edu/~liu32/data.html}. 
The locations are pictured as the red dots in Figure \ref{fig:cliamte} according to their longitudes and latitudes; the complete list of 125 location coordinates and 17 meteorological factors are collected in Section \ref{appsec:data-info-climate} of the Online Appendix.
In this application, data from each year serves as an observation, which is a {$125\times 17\times 12$} tensor. This dataset has been employed in \cite{chen2019graph} for estimating the support of the precision matrices, and in \cite{lozano2009spatial} for extreme value modeling. Following the spirit of \cite{chen2019graph}, the scientific interest in this section is to detect partial correlation (conditional dependence) alteration of locations {(i.e., $\k=1$)} before and after the \emph{1994 North American cold wave} in January 1994 \citep{agreement7january}. {Similarly, we apply the lag-one pre-processing step as in Section \ref{subsec:real-data-corr} to make the samples independent.} 
That is, the first group consists of the {lag-one differential tensor observations from 1991 to 1993, and the second group consists of those from 1994 to 1996}.
In contrast, \cite{chen2019graph} merges the monthly observations into an annual dataset {and considers a one-sample matrix-valued inference with dimension $125\times 17$.} 

Same as Section \ref{subsec:real-data-corr}, we first verify the separability by  \cite{aston2017tests} and the assumption is affirmed.
Next, we apply \texttt{T-BH}, \texttt{T-GAP}, \texttt{CL-SERA}, \texttt{LX-SERA} and the proposed \texttt{T-SERA} to the dataset. These five methods yield a rejection of {939, 958, 1, 0, and 1303} respectively, out of a total of 7750 hypotheses. Besides, the altered partial correlations found by \texttt{T-SERA} cover all those found by \texttt{T-BH} and \texttt{T-GAP}.
In addition, the tests by \cite{chen2019graph} and \cite{lyu2019tensor} suffer from trivial power, which may due to the over-correction of variances in their procedures (and thereby a decreasing signal-noise ratio) when the underlying nuisance covariances are dense.
In summary, it again suggests that the proposed \texttt{T-SERA} achieves the best performance. 
Finally, we visualize in Figure \ref{fig:cliamte} the {top 25} altered partial correlations found by \texttt{T-SERA} according to its weighted $p$-values. It is observed that the altered partial correlations are mostly detected in Midwestern regions of US. 
Such phenomenon is consistent with the historical fact that the extreme weather in 1994 mainly occurred in the Midwestern US \citep{schmidlin1997recent}.

\newpage
\bibliographystyle{apalike}
\bibliography{ref-sera}

\end{document}